\documentstyle[12pt,epsf]{article} \makeatletter
\@addtoreset{equation}{section} \makeatother

\newcommand{\st}[1]{\tilde{{\mbox{\boldmath $#1$}}}}
\newcommand{\tb}[1]{{\mbox{\boldmath $#1$}}}
\newcommand{\scr}[1] {\mbox{\scriptsize #1}}
\newcommand{\tscr}[1] {\mbox{\tiny #1}}
\newcommand{\mq} {m_{\scr{q}}}
\newcommand{\tc} {T_{\scr{c}}}
\newcommand{\bc} {\beta_{\scr{c}}}

\newcommand{\arcsinh} {\mbox{arcsinh}}
 \def\ie{{\sl i.e.\/}} 
\def\PR{{\sl Phys.\ Rev.\/}} \def\PRL{{\sl Phys.\ Rev.\ Lett.\/}}
\def\NP{{\sl Nucl.\ Phys.\/}} \def\PL{{\sl Phys.\ Lett.\/}}
\begin{document}
\thispagestyle{empty}
%
 \mbox{} \hspace{1.0cm}
 April 1994 
         \hfill HLRZ 54/93\hspace{1.0cm}\\
 \mbox{} \hfill BI-TP 93/76\hspace{1.0cm}\\
 \mbox{} \hfill TIFR/TH/94-12\hspace{1.0cm}\\
\begin{center}
\vspace*{1.0cm}
{{\large Spatial and Temporal Hadron Correlators below and above \\
         the Chiral Phase Transition\\}}
\vspace*{1.0cm}
{\large G. Boyd$^1$, Sourendu Gupta$^{1,3}$ \\
        F.~Karsch$^{1,2}$ and E.~Laermann$^2$} \\
\vspace*{1.0cm}
${}^1$ HLRZ, c/o Forschungszentrum J\"ulich, D-52425 J\"ulich, Germany\\
${}^2$ {Fak. f. Physik, Univ. Bielefeld, Postfach 100131, 
D-33501 Bielefeld, Germany}\\
${}^3$ Present Address: TIFR, Homi Bhabha Road, Bombay 400005, India\\
\vspace*{2cm}
{\large \bf Abstract}
\end{center}
\setlength{\baselineskip}{1.2\baselineskip}

Hadronic correlation functions at finite temperature in QCD, with four flavours
of dynamical quarks, have been analyzed both above and below the chiral
symmetry restoration temperature.  We have used both point and extended sources
for spatial as well as temporal correlators. The effect of periodic temporal
boundary conditions for the valence quarks on the spatial meson correlators has
also been investigated.  All our results are consistent with the existence of
individual quarks at high temperatures. A measurement of the residual
interaction between the quarks is presented.  \newpage\setcounter{page}{1}

\section{Introduction}

The spatial correlation functions of operators with hadronic quantum numbers
yield screening masses for static excitations in the QCD plasma
\cite{DeTar}. These give information on the physical excitations and
interactions in QCD at finite temperatures.  Chiral symmetry restoration at
temperatures $T>\tc$ is signalled by the degeneracy of screening lengths
obtained from pairs of opposite parity channels~\cite{DeTar,others,MTC}. At
such high temperatures, the vector (V), as well as the pseudovector (PV),
screening mass is nearly twice the lowest Matsubara frequency, $\Omega=\pi T$,
whereas the baryon screening mass is close to thrice this value. This indicates
that correlations in these channels are mediated, respectively, by the exchange
of two and three weakly interacting quarks \cite{MTC}.

Close to $\tc$, the scalar (S) and the pseudoscalar (PS) `meson' screening
mass is significantly smaller than the V/PV screening mass, but approaches
the latter when the temperature is raised to $2\tc$~\cite{gupta}. Thus, at
high temperature, this channel is also correlated through the exchange of two
weakly interacting quarks. Nevertheless, close to $\tc$, such an interpretation
does not seem to hold.

Since the lowest mass hadron is unlikely to change from a mesonic collective
state to a quark-like quasi-particle at a non-critical point, a pleasantly
consistent picture of the excitation spectrum would be obtained if, above
$\tc$, the S/PS correlations could be shown to be due to the propagation of two
unbound quarks, possibly still strongly interacting close to $\tc$. A means of
checking this was suggested and applied to quenched QCD~\cite{gupta}. This method
also yielded measurements of effective couplings in different spin channels. It
was seen that such a coupling is indeed larger in the S/PS channel than in the
V/PV channel.  In this paper we extend such studies to QCD with four flavours
of dynamical staggered quarks.

Hadronic correlators, and similarly the spatial Wilson loops \cite{borgs} and
spatial four point functions with point-split sources \cite{bernard}, should
provide information on the non-perturbative long distance structure of the
quark gluon plasma.  While the spatial and temporal quark propagators yield
identical information about quark screening masses~\cite{boyd}, this seems to
be in conflict with recent studies of temporal and spatial hadron
propagators~\cite{stamatescu}. In the latter case larger screening masses were
extracted from temporal than from spatial correlators, although one would have
expected to find smaller values as the lowest Matsubara mode is zero in this
case.  The determination of low momentum excitations from temporal propagators
at finite temperature is, however, difficult, due to the short ``time'' extent
of the lattice ($0<t<1/T$). This leads to a superposition of many high momentum
excitations. We will try to clarify this situation here by projecting onto low
momentum modes, by using wall source operators.

Furthermore, we will study the sensitivity of the extracted spatial screening
lengths to changes of the temporal quark boundary conditions. While bosonic modes
(bound states in mesonic operators) will not be sensitive to changes of the
usual anti-periodic boundary conditions to periodic ones, fermionic modes
(freely propagating quarks) will be sensitive to these changes. We will also
extend the studies of scalar and vector channel couplings~\cite{gupta} to the
case of QCD with dynamical fermions. 

The conclusion from these investigations
is that temporal and spatial propagators yield identical information on the
spectrum of finite temperature QCD below and above the chiral phase transition.
The behaviour of the (pseudo-)scalar screening length suggests that there are
no bound states in this quantum number channel above $\tc$.

This paper is organized as follows. In section 2 we introduce the observables
studied, and describe their behaviour for free staggered fermions, \ie, in the
lowest order of perturbation theory. Details of our measurements and the
results are given in section 3. A summary and conclusions are presented in the
section 4.

\section{Spatial and Thermal Direction Correlators}

In this section we consider correlation functions between operators separated
either in the spatial or the thermal direction. They are constructed on
lattices of size $V=N_\tau\times N_\sigma^3$ with $N_\sigma>N_\tau$.  Unless
otherwise mentioned, we shall use the lattice spacing as the unit of length.
The shorter direction introduces the temperature, via $N_\tau=1/T$. Spatial and
thermal direction correlation functions with given external meson momentum $P_\mu$
are defined as
\begin{eqnarray}
 G_s^H(x_3,\st P) \;&=&\; {1\over N_\sigma^2 N_\tau}\sum_{\st x}
        \langle H(x_3,\st x) H^\dagger(0,\st 0)\rangle 
        \exp\left({i\st P\cdot\st x}\right),\nonumber\\ 
 G_t^H(x_0,\tb P) \;&=&\; {1\over N_\sigma^3}\sum_{\tb x}
        \langle H(x_0,\tb x) H^\dagger(0,\tb 0)\rangle
        \exp\left({i\tb P\cdot\tb x}\right).
\label{corr}
\end{eqnarray}
For any 4-momentum $P_\mu$, we shall use the notation $\st P$ to denote the
3-momentum $(P_0,P_1,P_2)$, and $\tb P$ for $(P_1,P_2,P_3)$. Here $H$ denotes
an operator carrying mesonic quantum numbers. By an abuse of language, we shall
call $G_s$ and $G_t$ `meson correlators'. The name is not supposed to carry any
prejudice about the existence or absence of mesons in the theory.
 
In perturbation theory, these correlation functions can be described in terms
of the exchange of two (interacting) fermions. We set out our notation and
introduce some features typical of non-interacting staggered fermions below.
This corresponds to the leading order, ${\cal O}(g^0)$, of perturbation theory.
In section 3 the behaviour of the measured correlation functions will be
compared with these features.

\subsection{Spectral representations}

Hadronic correlation functions at $T=0$ are used to extract the mass of the
lowest excitation with given quantum numbers. A spectral representation of
correlation functions with quantum numbers $H$, in the form
\begin{equation}
  G^H(t)\;=\;\sum_i A_i {\rm e}^{-m_H^i t},
\label{spec}
\end{equation}
can be used to show that this necessitates the measurement of $G^H(t)$ at
Euclidean time separations much larger than the splitting between the lowest
and the first excited states, \ie, $t\gg 1/(m_H^1-m_H^0)$.  In the absence of
any {\em a priori} knowledge of this splitting, it is necessary to take as
large a time separation as possible, and to perform checks on the estimate of
$m_H^0$ so obtained. All this is well known.

At finite temperatures the physical extent of the thermal direction, $\tau$, is
bounded in physical units to $\tau\le1/T$. Due to the periodicity (or
antiperiodicity) imposed to obtain the thermal ensemble, correlation functions
can be followed only to $\tau\le1/(2T)$. In the general case this may not allow
the extraction of the lowest lying state in the manner discussed above. One
then has to use either a complicated ansatz for the correlation function,
involving several exponentials to approximate
eq.~(\ref{spec})~\cite{stamatescu}, or a more complicated operator which
projects out only the low momentum modes.

As the size of the spatial directions is not bounded in any way, it has become
customary to study spatial rather than temporal correlation functions. The
spatial correlation functions are the static correlations in the equilibrium
system, and hence are also of direct physical relevance.

For the interpretation of the correlation functions measured in these
simulations, one must discuss the spectral density functions underlying the
correlations.  While building models of the spectral densities, it is useful to
keep certain constraints in mind. At finite temperatures the heat-bath provides
a preferred frame of reference. As a result, the momentum representation of
spectral functions is given in terms of $p_0$ and $|\tb{p}|$ separately, where
$\tb p=(p_1,p_2,p_3)$. Dynamical modes are defined by poles in the complex
$p_0$ plane; and the movement of such poles with $|\tb p|$ are the dispersion
relations. In examining the spectral representation of the thermal and spatial
direction correlators it becomes clear that different aspects of the spectral
function are important for each. However, information on the poles of spectral
functions can be extracted from either of these correlators.

The residue at a pole can also be used to extract effective couplings. These
are usually defined as the integrals of discontinuities over cuts in the
complex plane of either $p_0$ or $|\tb p|$.  Note that on any lattice the
spectral function does not have any cuts, but only a set of poles. However,
effective couplings can still be extracted via sums over the residues at the
poles. One can expects finite lattice spacing effects to come from those poles
which develop into a cut in the continuum limit.

\subsection{Point Sources}

In the staggered discretization, local operators for currents 
carrying mesonic quantum numbers are written
in terms of fermion fields, $\chi(x)$, as
\begin{equation}
H(x) =  \phi_H( \tb x) \overline{\chi}(x_0, \tb x) \chi(x_0, \tb x).
\label{operator}\end{equation}
The phase factors $\phi_{H}$ project onto definite quantum
numbers~\cite{golt}, and are listed in Table~\ref{tab:phases}, for the
channels analyzed in this study.

\begin{table}[b]\leavevmode\begin{center}
\caption{Phases for the meson operators used for $G_t^H$. For $G_s^H$, $x_3$
         should be replaced by $x_0$. Also listed are kinematic factors 
         appearing in the ${\cal O}(g^0)$ perturbative calculations of the meson
         correlation functions in eqs. (\protect\ref{Gsdef}) and
         (\protect\ref{Gtdef}).} 
\vspace{1.5em}
\begin{tabular}{|c|c|c|c|}
\hline     
&~&~&~~\\[-0.3cm] 
Channel& $\phi_{H}(\tb x)$ & $f^H_{\scr{even}}$ & $f^{H}_{\scr{odd}}$  \\[0.1cm]
\hline
S &$1$                                           &$m^2-\omega\omega'$ &1\\
PS&$(-1)^{x_1+x_2+x_3}$                          &$m^2+\omega\omega'$ &1\\
PV&$(-1)^{x_1}+(-1)^{x_2}+(-1)^{x_3}$            &$3m^2-\omega\omega'$&3\\
V &$(-1)^{x_1+x_2}+(-1)^{x_2+x_3}+(-1)^{x_1+x_3}$&$3m^2+\omega\omega'$&3\\ 
\hline
\end{tabular}
\label{tab:phases}
\end{center}
\end{table}

Since the meson operators, $H(x)$, are composed of fermion operators,
it is clear that even for a fixed external meson momentum, $\st P$ or $\tb P$,
quarks with a spectrum of internal momenta will contribute to the correlation
functions $G_s^H$ and $G_t^H$. The structure of the hadronic correlation
functions is, therefore, more complicated than that of the quark correlator
\cite{boyd}.

This is clear when the correlation functions at ${\cal O}(g^0)$ are explicitly
written out. For the spatial correlation function the result is well known:
\begin{equation}
  G_s^H(x_3,\st P)/T^3\;=\;\frac{24N_\tau^3}{N_{\tau}N_\sigma^2} \sum_{\st p}
  f^{H} G^{q}_{s}(x_3,\st p) G^{q}_{s}(x_3,\st p') .
\label{Gsdef}
\end{equation}
The factors $f^{H}$ for even and odd sites are given in Table~\ref{tab:phases}.
The quark propagator $G^q_s$ is given by:
\begin{equation}
G_s^q(x_3,\st p)\;=\; \left\{
   \begin{array}{lr}
    {\displaystyle
    \frac{\sinh[E_s(x_3-N_\sigma/2)]}
    {\sinh(E_sN_\sigma/2)\cosh E_s}}
       &  (x_3 {\mbox{ odd}})\\
    {\displaystyle \rule{0em}{6ex} 
   \frac{\cosh[E_s(x_3-N_\sigma/2)]}
    {\sinh(E_sN_\sigma/2)\sinh 2E_s}}
       &  (x_3 {\mbox{ even}})\\
   \end{array}        \right.
\label{gsdef}
\end{equation}
The temporal correlation function is very similar:
\begin{equation}
G_t^H(x_0,\tb P)/T^3\;=\;\frac{24N_\tau^3}{N_\sigma^3} \sum_{\tb p}
   f^{H} G_t^q(x_0,\tb p) G_t^q(x_0,\tb p') , 
\label{Gtdef}
\end{equation}
with the quark propagator $G_t^q$ given by
\begin{equation}
G_t^q(x_0,\tb p)\;=\; \left\{
   \begin{array}{lr}
    {\displaystyle \frac{\cosh[E_t(x_0-N_\tau/2)]}
    {\cosh(E_tN_\tau/2)\cosh E_t}}
       &  (x_0 {\mbox{ odd}})\\
    {\displaystyle \rule{0em}{6ex} 
       \frac{\textstyle\sinh[E_t(x_0-N_\tau/2)]}
    {\textstyle\cosh(E_tN_\tau/2)\sinh 2E_t}}
       &  (x_0 {\mbox{ even}})\\
   \end{array}        \right.
\label{gtdef}\end{equation}

The sums in eqns. (\ref{Gsdef}) and (\ref{Gtdef}) are over internal quark
momenta.  These run over $p_0=(2n+1)\pi T$ and $p_i=2n\pi T$, with $i=1$, 2, 3
and $n=0,\pm1,\cdots$~. The quark and antiquark momenta, $p$ and $p'$
respectively, sum to give the meson momentum $P$, \ie ~$P_\mu=p_\mu+p'_\mu$. We
have used the abbreviation
\begin{equation}
E_{t,s}\;=\;\arcsinh(\omega_{t,s}),
\label{energy}
\end{equation}
along with the definitions
\begin{equation}
\omega_t^2=\mq^2+\sum_{k=1}^3\sin^{2}(p_{k})
\qquad{\mbox{ and}}\qquad
\omega_s^2=\mq^2+\sum_{k=0}^2  \sin^2(p_k).
\label{modes}\end{equation}
The variables $E'$ and $\omega'$ are obtained by replacing $p$ by $p'$, and 
$\mq$ is the bare quark mass.

\begin{figure}[htb]\leavevmode
\centerline{\epsfbox{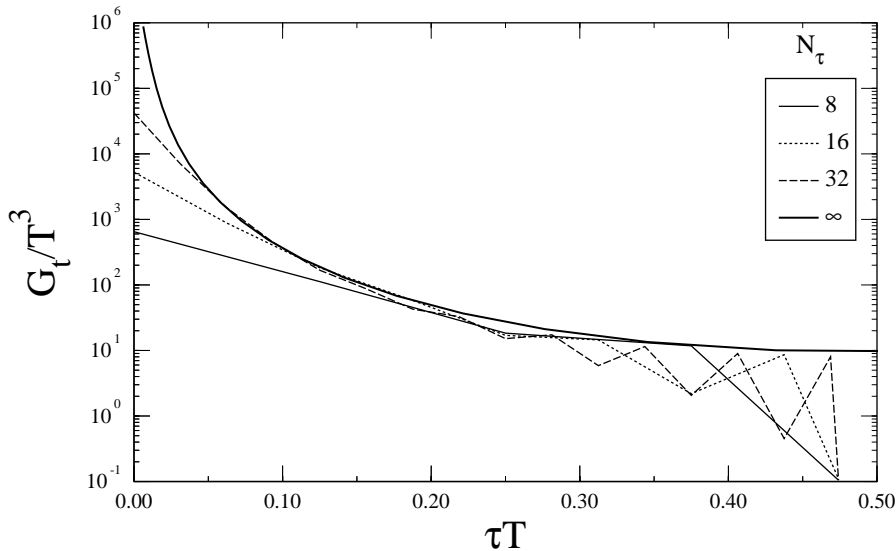}}
\caption{Thermal direction correlation functions at ${\cal O}(g^0)$ on
  lattices with $N_\sigma\to\infty$ and various $N_\tau$ values. The quark mass
  is $\mq=0.01$, and we define $\tau T=x_{0}/N_{\tau}$. The result for
  $N_\tau\to\infty$ is obtained from eq.~(\protect\ref{fmc}).  }
\label{corrfig}
\end{figure}

Recall that the spectrum of $p_0$ starts from the Matsubara frequency,
$\Omega=\pi T$, and reflects the antiperiodic boundary conditions usually
imposed in the thermal direction on the fermion fields when inverting the
Dirac equation. All other momentum components start from  zero. As a
result, the screening mass at $O(g^{0})$ is independent of the meson operator
$H$ and given by \cite{MTC} 
\begin{equation} 
\mu_H\;=\;2E_{\scr{s}}^{\tscr{min}}\;=\;
2\arcsinh\left({\mq^2+\sin^2\Omega}\right).
\label{mtcres}
\end{equation}

The decay of $G_s^H$ at $x_3\gg 1/T$ is controlled by $\mu_H$. At shorter
distances the higher modes also contribute and lead to a more rapid
decrease of the correlation function. At very short distances, $x_3<1/(2T)$, all
terms in the sums in eq. (\ref{Gsdef}) turn out to be important.

The thermal direction correlation function, $G_t^H$, does not yield extra
information in perturbation theory. Since $x_0<1/(2T)$, all terms in the sum in
eq. (\ref{Gtdef}) turn out to be important, and the correlation function falls
very rapidly due to the contribution of terms with large $E_t$ and $E'_t$.
This is shown in Figure~\ref{corrfig}. Also shown in the same figure is the
result obtained letting first $N_\sigma$ and then $N_\tau\to\infty$ at fixed
$T$. The formul\ae{} in eqs.  (\ref{Gtdef}) and (\ref{gtdef}) then reduce to
the continuum expressions for the correlators constructed from non-interacting
quarks. For zero external momentum, $\tb P=0$, and vanishing quark masses,
these take on the simple form
\begin{equation}
G_t^H (\tau,\tb P=0)\;=\;{96T^3\over\pi^2}
\int_0^{\infty}dy{y^2\over\cosh^2y}\cosh^2\bigl(y(2T\tau - 1)\bigr)
\label{fmc}\end{equation}
where $y$, $\tau$ and $T$ are related to the lattice variables via
$\tau=x_0a$, $\tau T=x_0/N_\tau$ and $y=E_t/(2T)$. Notice that, at
short distances, one obtains $G_t^H(\tau,\tb P=0)\sim\tau^{-3}$
from eq.~(\ref{fmc}). (Similar conclusions have been reached in the
continuum,~\cite{fri}.) 

Although the temporal direction is too short to single out the lowest
excitation for $T>0$, it is still instructive to study the behaviour of local
masses, $m(x_0)$, obtained from $G_t^H$.  These are obtained by comparing the
correlation functions on successive even (or odd) lattice sites, $x_0\pm1$. One
then solves
\begin{equation}
{G_t^H(x_0-1,\tb0)\over G_t^H(x_0+1,\tb0)}\;=\;
 {\cosh\left[m(x_0)(x_0-1-N_\tau/2)\right]\over
  \cosh\left[m(x_0)(x_0+1-N_\tau/2)\right]}\,,
\label{localmass}\end{equation}
for $m(x_0)$. The left hand side is obtained from either a measured or
perturbatively calculated correlation function. One then assumes on the right
hand side that the correlation function can be described by an ordinary meson
correlator consisting of a single cosh. (For the simple case of non-interacting
quarks we know, of course, that we should rather use the sum over squares of
quark correlators applicable to the given source and sink.)  A feeling for this
local mass may be obtained by noting that in the limit $N_\tau\to\infty$, at
fixed $T$ (\ie, $a\to0$), eq. (\ref{fmc}) can be used to show that
\begin{equation}
m(x_0)x_0 = m(\tau)\tau\;\simeq\;3 \qquad(N_\tau\gg x_0\gg 0).
\label{asymptotic}\end{equation}

\begin{figure}[bt]\leavevmode
\centerline{\epsfbox{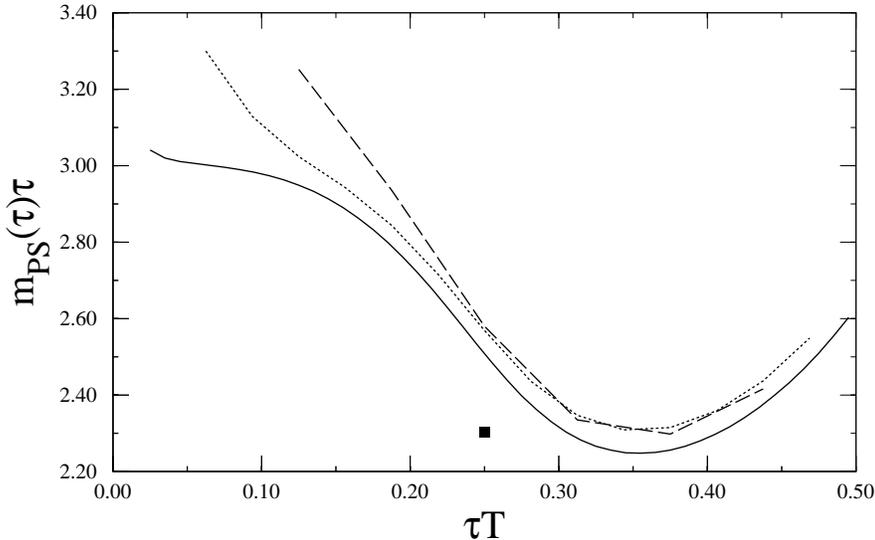}}
\caption{Scaled local masses at $O(g^{0})$, $m(\tau)\tau$,  versus the 
  scaled separation $\tau T=x_0/N_\tau$, obtained from eqs.
  (\protect\ref{Gtdef}) and (\protect\ref{gtdef}) for $N_\sigma\to\infty$ for
  $N_\tau=16$ (dashed line) and 32 (dotted line). The full line shows the
  result for $N_\tau=\infty$ (eq.  (\protect\ref{fmc})). The filled square shows
  the result for $m(\tau=0.25/T)$ on an $8\times16^3$ lattice.  }
\label{lmfig}
\end{figure}

Analytic results for the local masses at ${\cal O}(g^0)$ in perturbation theory
are shown in Figure~\ref{lmfig}. Note that, in practice, $m(\tau)\tau$ remains
close to 3 for all values of $\tau T$.  The rise of the local meson mass at
short distances, when extracted in this fashion, reflects the increasing
importance of higher momentum excitations. We note that even for the largest
possible separation $x_0=N_{\tau}/2$, or equivalently $\tau=1/(2 T)$, the local
mass remains large in units of the temperature, $m/T\approx 6$.  The result for $\tau=
1/(4T)$, obtained on an $8 \times 16^3$ lattice, is also shown in the figure.
This is the only value of $\tau$ which we will be able to study numerically.

As a concluding remark to this section, let us point out a distinctive feature
of a meson propagator consisting of non-interacting quarks, $G_t^H(x_0)$,
namely the strong oscillation between even and odd sites visible in Figure
\ref{corrfig}.  This is seen for any non-zero lattice spacing (finite
$N_\tau$), and persists in higher orders of perturbation theory.  Such
oscillatory behaviour is also seen in the spatial correlation functions in
perturbation theory.  In the low-temperature phase of QCD, however, where one
has a genuine bound state, the correlator does not show such a behaviour.

\subsection{Wall Sources}

From the previous discussion we conclude that the analysis of the correlation
functions (and especially the temporal correlation functions) between local
hadronic operators is complicated, owing to the momentum sums in eqs.
(\ref{Gsdef}) and (\ref{Gtdef}). A determination of the low momentum
excitations through local masses or fits with a few exponentials seems to be
difficult, as all higher energy levels, coming from higher quark momenta,
contribute.

The contributions from higher quark momenta can be suppressed by the judicious
choice of a wall source, which creates quarks with only the lowest allowed
momentum. For correlations measured in the thermal direction, the boundary
conditions on the transverse slice are periodic. In this case one can project
onto zero quark momentum via the meson sources
\begin{equation} H_{\scr{wall}}(x_0,\tb{x}) = 
  {8\over N_\sigma^3}
  \sum_{\stackrel{\tb{e},\tb{e'}}{\scr{even}}} 
  \phi_H( \tb x)
  \overline{\chi}(x_0, \tb{x}+\tb e) \chi(x_0, \tb{x}+\tb e').
\label{wallt}\end{equation}

For correlators measured in the spatial direction, the antiperiodic boundary
condition in the thermal direction means that one can not project onto zero
momentum. Instead a projection onto the Matsubara momentum $\st\Omega= (\pi
T,0,0)$ must be used:
\begin{equation}
  H_\Omega(x_3) =  {8\over N_\sigma^2 N_\tau}
  \sum_{\stackrel{\tb{e},\tb{e'}}{\scr{even}}} 
  \phi_H( \st x) \overline{\chi}(x_3, \st x +\st{e}) 
  \chi(x_3, \st x + \st{e}')
  \exp\left({i(\st{e}-\st{e}')\cdot\st\Omega}\right).
\label{wallx}\end{equation}
The correlation functions are measured between a wall source and a point sink,
with a sum over all sinks in the plane transverse to the direction of
propagation.  

The $O(g^0)$ perturbative calculation in eqs.~(\ref{Gsdef}) and~(\ref{Gtdef})
is simplified if the only contribution is from quarks with momentum
$\tb{p}=\tb{0}$ or $\st{p}=\st{\Omega}$.  For example, the temporal correlator
becomes the square of a single quark propagator,
\begin{equation}
  G_{t}^{H,\scr{wall}}(x_0,\tb 0)/T^3\;=\;3(N_\tau^3/N_\sigma^3) 
  f^{H} G_t^q(x_0,\tb 0) G_t^q(x_0,\tb 0).
\label{GtWdef}
\end{equation}

The biggest gain in using extended sources is obtained for the thermal
direction correlation functions. At ${\cal O}(g^0)$ in perturbation theory the
masses  are then
given by 
\begin{equation}
m^{\scr{wall}}\;=\; 2\arcsinh\left(\mq\right).
\label{mwall}\end{equation}
Local masses can then be extracted by using eq.~(\ref{localmass}).
We note that at ${\cal O}(g^0)$ the meson correlator now has the form
$\cosh^{2}(\mq(x_0-N_\tau/2))$, which is well fitted by a
$\cosh(2\mq(x_0-N_\tau/2))$ over most of the temporal range.

At ${\cal O}(g^0)$ in perturbation theory the local masses thus measure the
mass of the lowest excitation.  At higher orders this is modified due to
diagrams which can be separated into thermal corrections to each of the quark
propagators and interactions between the propagating quark and antiquark. Each
of these effects begins at ${\cal O}(gT)$.  If the effective interaction
between the two quarks can be neglected, as we expect for the V channel, then
the effect of interactions can be wholly subsumed into replacing the quark mass
$\mq$ in eq. (\ref{mwall}) by the thermal quark mass, $\mq^{\scr{eff}}$. This
was extracted from the temporal quark correlator in~\cite{boyd} using the same
configurations used here.  It is not unrealistic to hope that such a result may
hold beyond perturbation theory as well.

\subsection{Varying Boundary Conditions}

In our attempt to clarify the nature of the excitations in the plasma we have
also varied the temporal boundary condition on the valence fermion fields from
which the meson operators are constructed. The boundary conditions of the sea
quarks, driving the generation of the gauge field ensemble, have been left
untouched.  This should give a direct view of the nature of excitations in the
plasma.

If the lowest excitation in a channel of fixed quantum numbers consists
of a meson, then the spectral representation contains a pole. Thus, we
are assured that this state will be seen in the correlation of generic
operators with these quantum numbers. In particular, whether we use periodic
or anti-periodic boundary conditions on the fermion source, the lowest mass
obtained out of the correlation functions must be the same. The correlation
functions should thus show little dependence on the choice of boundary
conditions. 

However, if a given quantum number can only be obtained by exchanging more than
one particle, then the spectral representation contains a cut.  If the
exchanged particles are fermions,
then the contribution to the correlation function depends on the
boundary conditions. In fact, at ${\cal O}(g^0)$, it is easy to see that eqs.
(\ref{Gsdef}) and (\ref{gsdef}), which define the spatial correlation
functions, depend on the temporal boundary conditions via the spectrum of $p_0$
alone. For periodic boundary conditions in the thermal direction, the allowed
values are $p_0=2n\pi T$ with $n=\{0,1,\cdots\}$, which yields a screening mass
equal to
\begin{equation}
\mu\;=\;2E_s^{\scr{min}}\;=\;2\arcsinh(\mq).
\label{pbc}\end{equation}
The qualitative change in going from anti-periodic to periodic boundary
conditions in the thermal direction is thus expected to be significant, and
should show up in the correlation functions. Interactions will modify this in
the same way as discussed above.

\subsection{Effective inter-quark Couplings}

One may use the meson propagators to calculate an effective four fermion
coupling as used in, for example, the Nambu--Jona-Lasinio
model~\cite{klimt}.  The propagator at zero four-momentum is equal to the
generalised susceptibility~\cite{gupta}, 
\begin{equation} 
\chi_{\pi,\rho} =
\tilde{G}_{\pi,\rho}(0).  
\end{equation}

Letting $\tilde{G}_{\pi,\rho}^{0}(0) = \chi_{\pi,\rho}^{0}$ be the
susceptibility
for non-interacting quarks in the pion and rho channels respectively, 
one obtains from the Dyson equation the
result:
\begin{equation}
\chi_{\pi,\rho} = 
    \frac{\chi_{\pi,\rho}^{0}}{1-g_{\pi,\rho}\chi_{\pi,\rho}^{0}} ,
\end{equation}
where $g_{\pi,\rho}/2$ is the four fermion coupling in the effective Lagrangian,
following the convention of~\cite{klimt}.

Solving for $g_{\pi,\rho}$ one obtains:
\begin{equation}
g_{\pi,\rho}T^{2} = \frac{1}{N_{\tau}^{2}}\left(
                   \frac{1}{\chi_{\pi,\rho}^{0}} - \frac{1}{\chi_{\pi,\rho}}
                                          \right),
\label{eq:coupl}
\end{equation}
where the susceptibility extrapolated to zero quark mass is used.
\footnote{The corresponding formula, eq. 19, given in~\protect\cite{gupta},
  contains a minor error in the normalization.}

\section{Results}

In this section we describe the results of our measurements on configurations
generated by the $M\tc$ collaboration on lattices of size $8\times 16^{3}$ with
4 flavours of dynamical fermions having a bare mass $\mq a=0.01$. Recall that
the phase transition was observed at a coupling of $\beta=5.15(5)$~\cite{mtca}.
At this coupling we have two sets of configurations. The set labelled B
corresponds to a run which started in the low-temperature (chiral symmetry
broken) phase and remained there. The set labelled S was started in the
high-temperature (chirally symmetric) phase and did not tunnel into the other
phase.  The couplings $\beta=5.1$ and $5.2$ cover the temperature range $T/\tc
= 1\pm 0.2.$

Results are presented for both point sources and wall sources. For the wall
sources we have fixed the configurations to the Coulomb gauge in the hyperplane
containing the source, transverse to the direction of propagation.

We have around twenty configurations at each value of the coupling, with four
sources per configuration for point sources, and one for wall sources.  In
calculating the errors for the point sources we first blocked the four sources.
The errors quoted reflect the statistical fluctuations alone.

\subsection{Thermal direction point source correlators}

\begin{figure}[htb]\leavevmode
    \hspace{-8mm}\epsfbox[0 0 135 246.5]{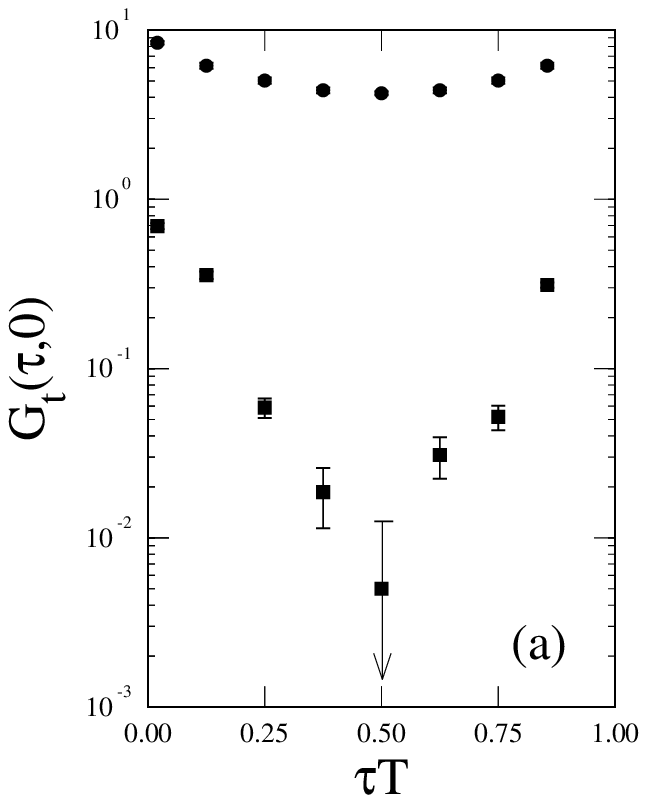}
    \epsfbox[0 0 135 246.5]{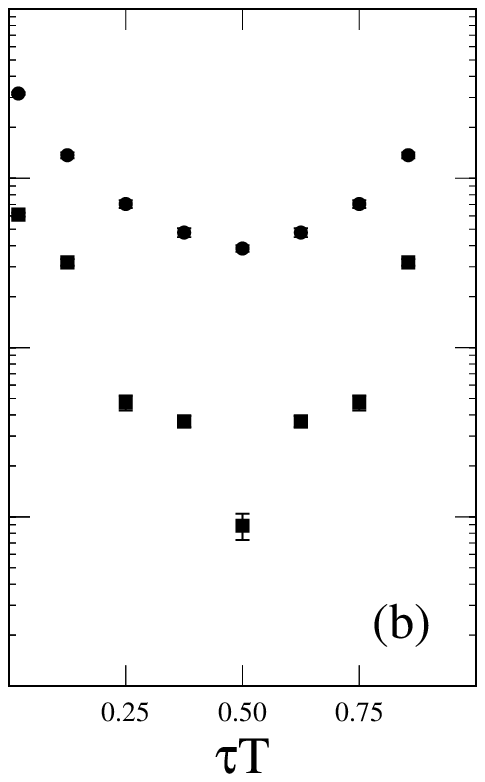}
    \epsfbox[0 0 135 246.5]{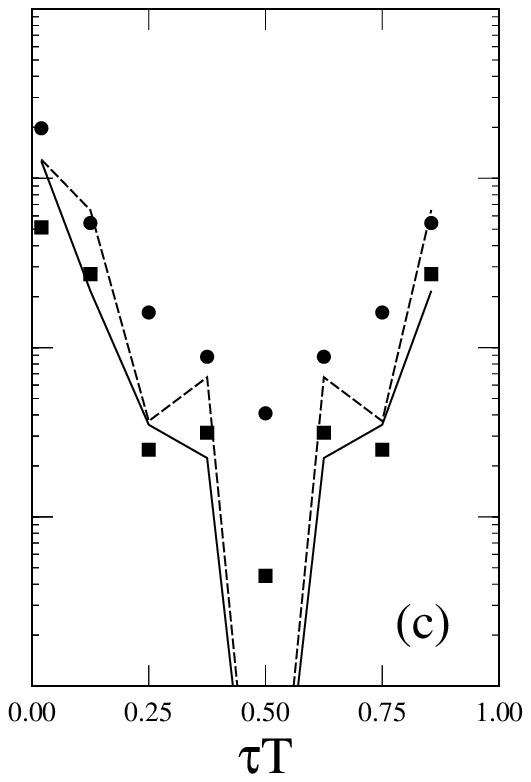}
  \caption{The thermal direction correlators for PS (filled circles) and V
    (filled squares) at $\beta=5.1$ (a), 5.3 (b) and 6.5 (c), shown as a
    function of $\tau T$. In (c) the lowest order results
    for PS (solid line) and V (dashed line) correlators are also shown.
    } 
\label{tempcor}
\end{figure}

The measured values of $G_t^H$ are shown in Figure~\ref{tempcor}. Note that the
pseudoscalar correlator for $T<\tc$ is very well described by a hyperbolic
cosine. A changeover to the oscillatory behaviour characteristic of free
fermions, as discussed in the previous section,
is visible in our thermal correlation functions as the critical
coupling, $\bc=5.15$, is crossed. This change in behaviour is most
noticeable when wall sources are used. Note also that the oscillation between
even and odd sites is enhanced in the vector channel. This is in accordance
with the perturbative calculation, and is due to the fact that $f^V\geq f^{PS}$
(see Table~\ref{tab:phases}).

\begin{figure}[tbp]\leavevmode
\centerline{\epsfbox[0 0 387 240]{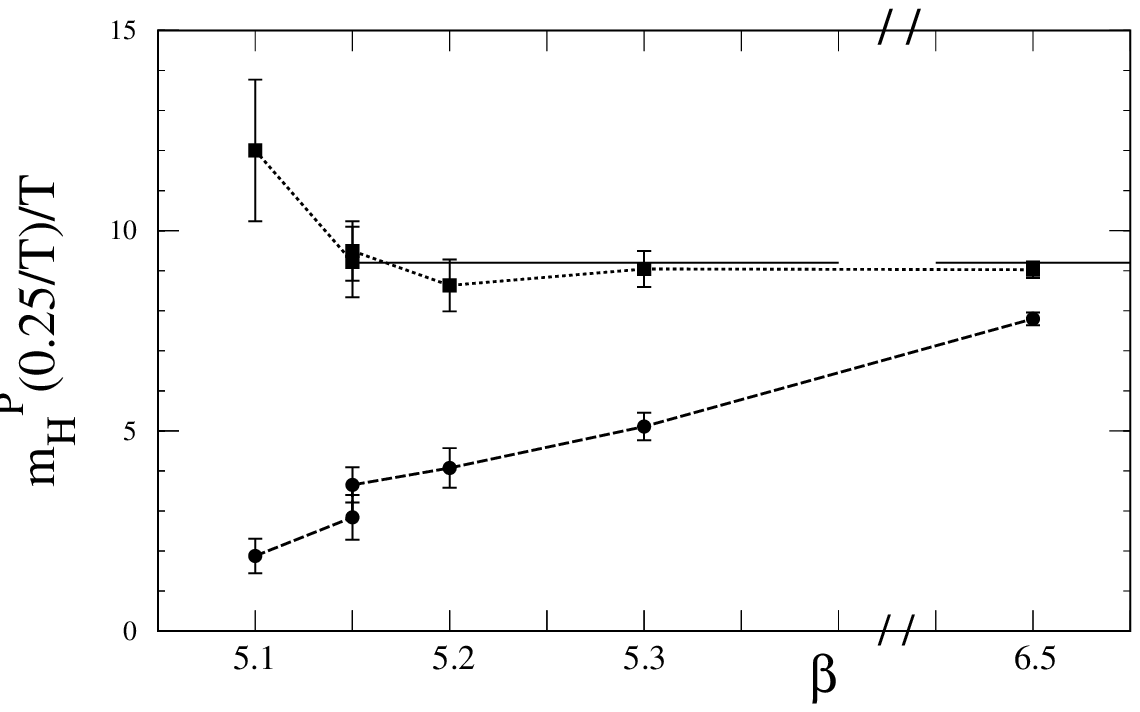}}
\caption{Local masses $m_{\scr{H}}^{\scr{P (=point)}}$ at 
  distance $\tau T=1/4$ from thermal direction V (squares) and PS (circles)
  correlation functions using point sources, shown as a function of $\beta$.
  Also shown is the corresponding result computed in a theory of
  non-interacting fermions on an $8\times 16^3$ lattice at the same $\tau$
  (full line).  }
\label{localtemp}
\centerline{\epsfbox[0 0 387 240]{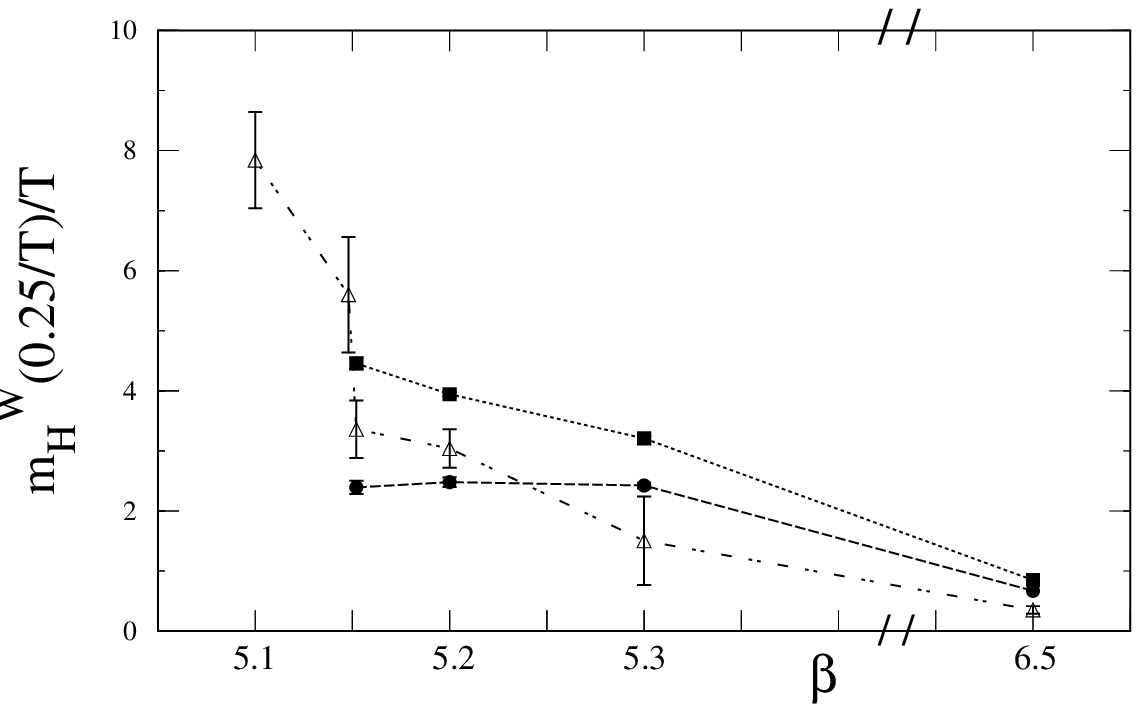}}
\caption{Local masses $m_{\scr{H}}^{\scr{W (=wall)}}$ from correlators 
  constructed with wall sources in the PS (circles) and V (squares) channels in
  the high temperature phase.  Also shown is $2m_q^{\scr{eff}}$ (triangles)
  from \protect\cite{boyd}.  Lines have been drawn to guide the eye.  }
\label{tempwall}
\end{figure}

In Figure~\ref{localtemp} we show the local masses at distance $\tau=0.25/T$
extracted from the V and PS correlation functions using eq.~(\ref{localmass}),
(and hence the implied ansatz,) in the thermal direction.  These are also
listed in Table~\ref{mresults}.  We find that the local mass in the V channel
is close to the perturbative value, $m_{V}=9.4/T$ (see Figure~\ref{lmfig}).
However, the local mass in the PS channel approaches this value rather slowly
with increasing temperature. Such a behaviour is very similar to that of the
screening masses extracted from spatial correlation functions.

\begin{table}[b]
\caption{Local masses from thermal direction V and PS correlation
    functions at distance $\tau=1/(4T)$. Results obtained from 
    point and wall source operators are shown. For comparison we also
    list twice the effective quark mass \protect\cite{boyd}.}
\begin{center}\begin{tabular}{|c|c|c|c|c|c|}
    \hline
    $\rule{0ex}{3ex}\beta$&$m_{PS}^{\scr{point}}$&$m_V^{\scr{point}}$
                              &$m_{PS}^{\scr{wall}}$&$m_V^{\scr{wall}}$
                                                    &$2\mq^{\scr{eff}}$\\
    \hline
    5.1     & 0.24(5)  & 1.5(2)   &             &           & 0.9(1)   \\ 
    5.15(B) & 0.36(7)  & 1.2(1)   &             &           & 0.6(1)   \\ 
    5.15(S) & 0.46(6)  & 1.19(9)  &  0.29(1)    & 0.56(1)   & 0.42(6)  \\ 
    5.2     & 0.51(6)  & 1.08(8)  &  0.31(1)    & 0.493(5)  & 0.38(4)  \\ 
    5.3     & 0.64(4)  & 1.13(6)  &  0.303(6)   & 0.401(5)  & 0.19(9)  \\ 
    6.5     & 0.98(2)  & 1.13(3)  &  0.0838(4)  & 0.106(2)  & 0.044(8) \\
    \hline
  \end{tabular}\end{center}
\label{mresults}
\end{table}

These local masses are similar in magnitude to the screening
masses. This is accidental. It is due to the fact that for $N_\tau=8,
N_\sigma=16$ we can only extract a local temporal mass at $\tau T=1/4$. In a
free fermion theory on this size lattice we find that the local mass at this
distance is about $10T$. This just happens to be close
to the meson screening mass in free fermion theory on this size lattice, $9T$.
When the lattice size is changed, this accidental concordance is removed (see
eq.~(\ref{asymptotic})).

\subsection{Thermal direction wall source correlators}

Local masses have similarly been extracted from the temporal wall source
correlators.  The results obtained for $m^{\scr{wall}}$ are also collected in
Table~\ref{mresults}. We note that these masses are indeed much smaller than
the local masses obtained using point sources. The projection onto the lowest
momentum excitation with these correlation functions thus seems to be rather
efficient.

The relation in eq. (\ref{mwall}) seems to hold, at least qualitatively,
for both the vector and the pseudo-scalar channels, when the effective quark
mass $\mq^{\scr{eff}}$ given in~\cite{boyd} is used for the quark mass.  The
masses in the PS channel close to $\bc$ are smaller than those in the
vector channel, indicating strong residual interactions between fermions. The
PS masses then approach the mass in the vector channel as the coupling
increases.

We also note that the vector mass remains larger than $2\mq^{\scr{eff}}$,
indicating the importance of the remaining inter-quark interactions. One may
try to parametrize these residual interactions in a potential model relating
the difference in the pion and rho masses to different spin--spin interactions
in these quantum number channels. The masses then receive contributions from
the effective quark masses, as well as the scalar ($E_{\scr{scalar}}$) and spin
dependent ($E_{\scr{spin}}$) part of the quark--anti-quark potential:
\begin{eqnarray}
  m_{PS} & = & 2\mq^{\scr{eff}} + E_{\scr{scalar}} - \frac{3}{4}E_{\scr{spin}} \\
  m_{V} & = & 2\mq^{\scr{eff}} + E_{\scr{scalar}} + \frac{1}{4}E_{\scr{spin}} .
  \label{eq:spin}
\end{eqnarray}
We find that the meson masses can then be parametrized by a scalar term that is
consistent with zero, and a spin term approximately equal to the effective
quark mass, and with the same temperature dependence.

\subsection{Anti-Periodic and Periodic Boundary Conditions}
 
\begin{figure}[htb]\leavevmode
 \centerline{\epsfbox[40 0 190 246.5]{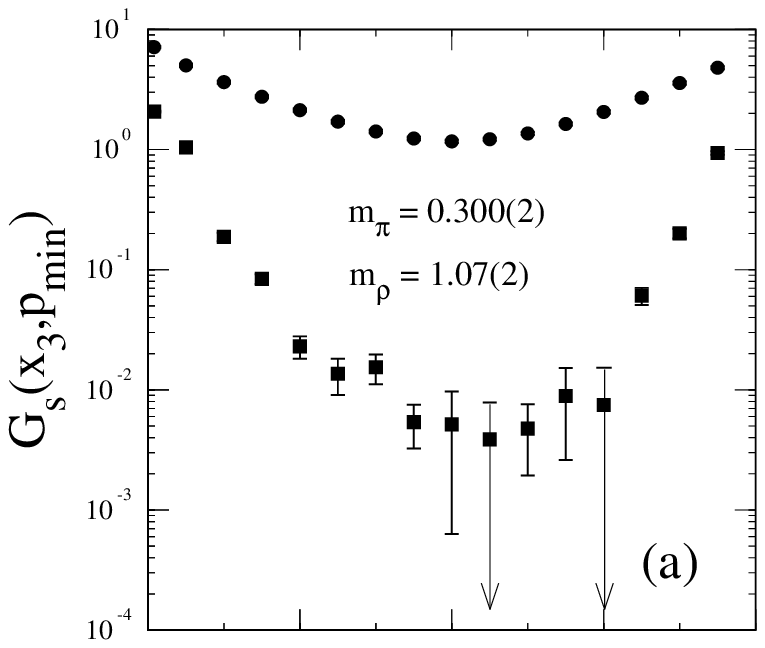}
             \epsfbox[0 0 190 246.5]{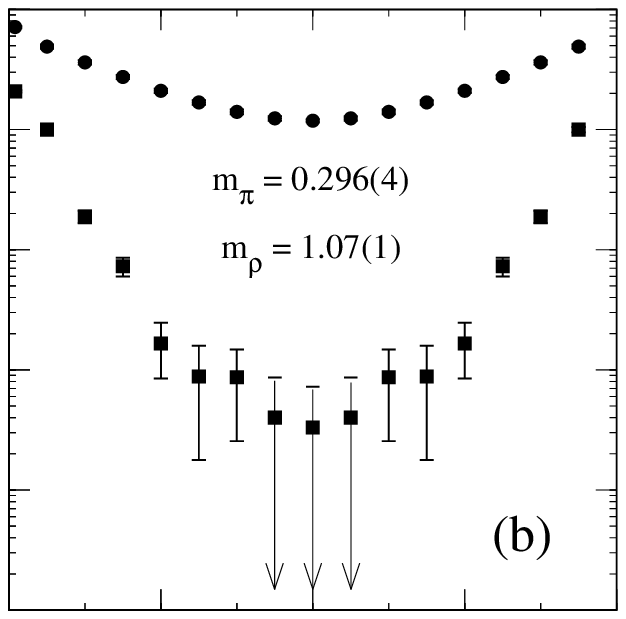}}
 \centerline{\epsfbox[40 0 190 185]{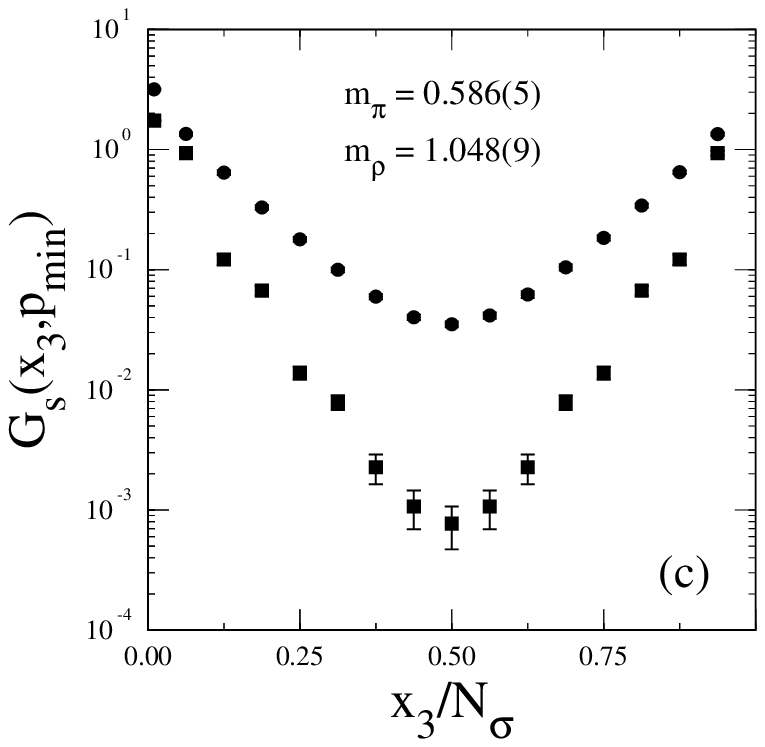}
             \epsfbox[0 0 190 185]{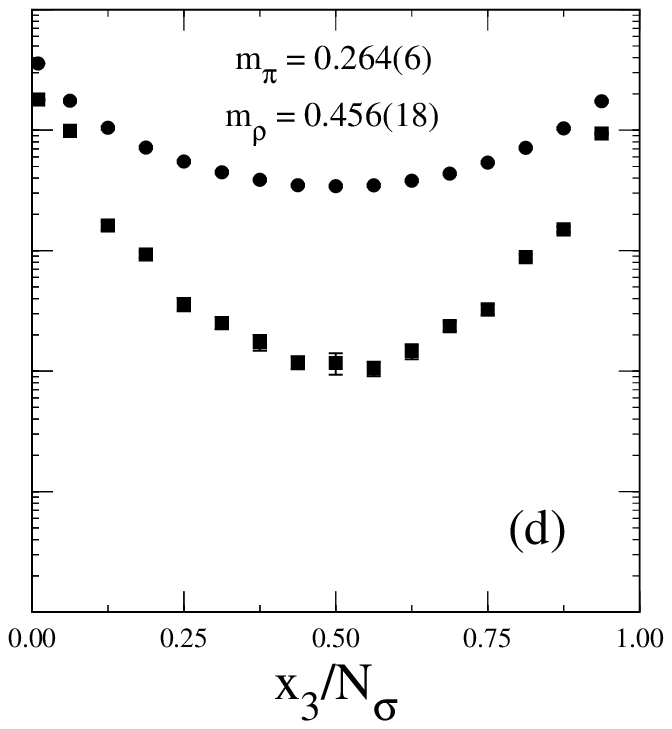}}
 \caption{$G_s^H$ in the PS (circles) and V (squares) channels. Figures (a) and
   (b) correspond to $\beta=5.1$, while figures (c) and
   (d) correspond to $\beta=5.3$.
   The valence quarks have anti-periodic boundary conditions in the thermal
   direction for (a) and (c),
   with (b) and (d) having periodic boundary conditions in the
          thermal direction. The masses shown were obtained from fits.
          }
 \label{cb}
\label{hb}
\end{figure}
In Figure~\ref{hb} (a) and (b) $G_s^{PS}$ and $G_s^V$ are shown for $T<\tc$ for
both periodic (per) (a) and anti-periodic (aper) (b) boundary conditions in the
thermal direction. The correlation functions are unaffected by this change, and
the screening masses, shown in Table~\ref{scrm}, do not change within errors.
Thus, these masses reflect bosonic poles in the spectral function.

In contrast, there is a remarkable difference between the correlation functions
obtained for periodic and anti-periodic boundary conditions for $T>\tc$, as
shown in Figure \ref{cb} (c) and (d). The correlation functions become much
flatter when the boundary conditions are periodic, and the screening mass drops
in both the PS and V channels. Varying the boundary conditions thus suggests
the absence of genuine mesons in the high temperature phase of QCD.

It can be seen that the vector screening mass $\mu_V^{\scr{per}}$ closely
satisfies the relation in eq.~(\ref{pbc}), when the quark mass is taken to be
the non-perturbative quark mass measured at the same coupling~\cite{boyd}.
Thus, the effects of interactions, in this angular momentum channel, can be
almost entirely lumped into the effective quark mass.  (This relation also
holds below the transition, and gives a definition of the constituent
quark mass.)  As is already known \cite{gupta}, this simplification does not
hold in the $PS$ channel, and an effective interaction between quarks remains.
The value of $\mu_{PS}^{\scr{per}}$ is accordingly somewhat smaller than
$\mu_V^{\scr{per}}$.
 
\begin{table}[b]
\caption{Screening masses in the PS and V channels for both anti-periodic (aper)
  and periodic (per) boundary conditions in the thermal direction. In the
  chirally symmetric phase results obtained with wall source operators (W) are
  also quoted. The screening mass expected for this size lattice, using
  non-interacting quarks, is $2 \pi T = 1.2$ for point sources and $2 \pi T =
  1.0$ for wall sources.  
  } 
\vspace{1ex}
\begin{center}
\begin{tabular}{|c|c|c|c|c|}
\hline
$\rule{0ex}{3ex}\beta$&$\mu_{PS}^{\scr{aper}}$&$\mu_{PS}^{\scr{per}}$
&$\mu_V^{\scr{aper}}$&$\mu_V^{\scr{per}}$\\
\hline 
5.1     & 0.300(2)   & 0.296(4)     & 1.07(2)   & 1.07(1)     \\
5.15(B) & 0.296(7)   & 0.301(5)     & 1.15(1)   & 0.84(5)     \\
5.15(S) & 0.394(6)   & 0.299(7)     & 0.947(7)  & 0.68(3)     \\
5.15(SW)&            & 0.270(3)     &           & 0.552(3)    \\
5.2     & 0.469(7)   & 0.294(8)     & 0.979(10) & 0.548(18)   \\
5.2(W)  &            & 0.259(3)     &           & 0.463(3)    \\
5.3     & 0.578(5)   & 0.264(6)     & 1.048(9)  & 0.456(18)   \\
5.3(W)  & 0.553(4)   & 0.218(5)     & 0.820(2)  & 0.337(6)    \\
6.5     & 0.899(3)   & 0.112(9)     & 1.101(3)  & 0.113(21)   \\
\hline
\end{tabular}\end{center}\label{scrm}\end{table}

\begin{figure}[htb]\leavevmode
 \centerline{\epsfbox{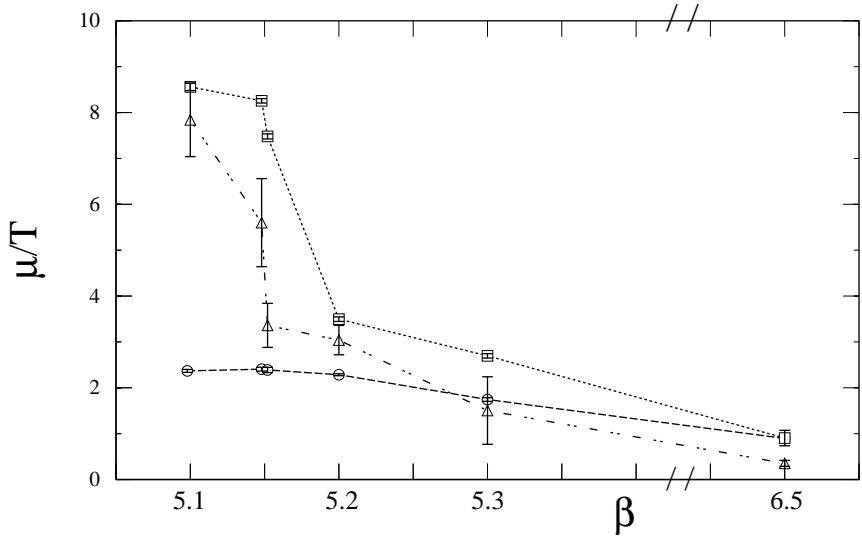}}
 \caption{Screening masses extracted from spatial correlation functions using 
   point sources with periodic temporal boundary conditions. Shown is the rho
   (squares) and pion (circles) screening mass, extracted in both cases from
   fits to the correlator on even sites.  Also shown is twice the effective
   quark mass (triangles) determined on the same gauge field configurations in
   the Landau gauge \protect\cite{boyd}. Lines are drawn to guide the eye.}
 \label{pbcwall}
\end{figure}

In Table~\ref{scrm} and Figure
\ref{pbcwall}  results for screening masses from the wall source operators
(eq.~(\ref{wallx})) are presented. These are a little smaller than those
obtained from point source operators, which indicates that the spatial
direction is not large enough to eliminate higher terms in the sum in eq.
(\ref{Gsdef}).  Presumably, slightly larger spatial lattices would be required
for this. Experience from~\cite{gupta} shows that spatial sizes
$N_\sigma\approx4N_\tau$ generally suffice to eliminate the effect of the
higher modes.

The values of $\mu_H^{\scr{per}}$ are similar to the values of local masses
extracted from wall source operators in the thermal direction. This is indeed
to be expected in perturbation theory. At ${\cal O}(g^0)$ these two masses
should be the same, while differences can arise at higher orders in $g$. Thus
in the limit $\mq\to0$, both these quantities are sensitive to $O(gT)$ thermal
corrections.  The screening masses in Figure \ref{pbcwall} may be compared with
the wall source results shown in Figure~\ref{tempwall}.


\subsection{Effective inter-quark Couplings}

The effective four fermion coupling, defined in eq.~(\ref{eq:coupl}), is
presented in table~\ref{tab:coupling}. Notice that the coupling in the pion
channel is about four times stronger than that in the rho channel, supporting
the hypothesis that the difference between the pion and rho screening masses
lies in different interaction strengths between the quarks in the two channels.
The numbers obtained at the transition are also comparable with those obtained
in the quenched approximation~\cite{gupta}.

\begin{table}[b]
\caption{Effective four fermion couplings for the pion $(g_{\pi})$
        and rho $(g_{\rho})$ in the limit of zero quark mass. The
        susceptibilities for non-interacting quarks are:
        $1/\chi^{0}_{\pi}=0.5478$ and $1/\chi^{0}_{\rho}=0.3680$.
        \label{tab:coupling}
}
\begin{center}
\begin{tabular}{|l|l|l|}
\hline
$\rule{0ex}{3ex}\beta$ & \multicolumn{1}{c|}{$g_{\pi}T^{2}$} 
        & \multicolumn{1}{c|}{$g_{\rho}T^{2}$} \\
\hline
5.15(S) & 0.00795(6) & 0.00184(10) \\
5.2     & 0.00785(8) & 0.00189(11) \\
5.3     & 0.0074(1)  & 0.00184(15) \\
6.5     & 0.005(8)   & 0.0010(4) \\
\hline
\end{tabular}
\end{center}
\end{table}

Taking the transition temperature to be $T_{c}=0.14$GeV~\cite{mtcm} the
following physical values for the couplings in the chirally symmetric phase at
the transition are obtained: $g_{\pi} = 0.41$GeV$^{-2}$ and $g_{\rho} =
0.094$GeV$^{-2}$.  These numbers may be compared with the values quoted
in~\cite{klimt}, obtained from fits to the experimental meson data at $T=0$:
$g_{\pi} = 4.90$GeV$^{-2}$ and $g_{\rho} = 3.25$GeV$^{-2}$. The couplings at
high temperature are well below the critical coupling at which the
Nambu--Jona-Lasinio model first shows chiral symmetry breaking, and provide a
further indication that neither channel has a low lying bound state.

\section{Conclusions}

The pion and rho correlators are, above the phase transition, sensitive to both
the boundary conditions and the type of source used.  This is not seen below
the phase transition. Since the changes in the correlator are of the form one
expects if unbound fermions play a direct role in the spectral function, this
provides evidence for the existence of a two fermion cut dominating the
spectral function.

Above the phase transition the screening masses in both the PS and vector
channels are consistent with twice the effective quark mass plus some residual
interactions. As a measure of this interaction, and hence as a summary of the
relevant physics of the system, we extracted an effective four fermion
coupling. This was four times stronger for the PS channel than it was for the
vector channel, and an order of magnitude smaller than the couplings used in
Nambu--Jona-Lasinio models at zero temperature.

The structure of correlators in the vector channel above $\tc$ generally agrees
quite well with the behaviour expected from leading order perturbation theory;
however, this is not the case for the pseudo-vector channel below $2\tc$. Here
the correlators and masses are seen to approach the perturbation limit rather
slowly. 

One is left with a consistent picture of a plasma phase consisting of
deconfined, but strongly interacting quarks and gluons in the temperature range
from $\tc$ to $2\tc$.

\end{document}